\documentclass[10pt, superscriptaddress,twocolumn]{revtex4}

\usepackage{upgreek,hyperref}
\usepackage{graphicx}

\begin{document}

\title{ Efficient photonic reformatting of celestial light for diffraction-limited spectroscopy}

\author{David G. MacLachlan}
\altaffiliation{Author contributed equally to this work}
\email{dgm4@hw.ac.uk}
\affiliation{SUPA, Institute of Photonics and Quantum Sciences, Heriot-Watt University, Edinburgh, EH14 4AS, UK}
\author{Robert J. Harris}
\altaffiliation{Author contributed equally to this work}
\email{rharris@lsw.uni-heidelberg.de}
\affiliation{Department of Physics, University of Durham, South Road, Durham, DH1 3LE, UK}
\author{ Itandehui Gris-S{\'a}nchez}
\affiliation{Department of Physics, University of Bath, Claverton Down, Bath BA2 7AY, UK}
\author{Timothy J. Morris}
\affiliation{Department of Physics, University of Durham, South Road, Durham, DH1 3LE, UK}
\author{Debaditya Choudhury}
\affiliation{SUPA, Institute of Photonics and Quantum Sciences, Heriot-Watt University, Edinburgh, EH14 4AS, UK}
\author{Eric Gendron}
\affiliation{LESIA, Observatoire de Paris, Meudon, 5 Place Jules Janssen, 92195 Meudon, France}
\author{Alastair G. Basden}
\affiliation{Department of Physics, University of Durham, South Road, Durham, DH1 3LE, UK}
\author{Izabela J. Spaleniak}
\affiliation{SUPA, Institute of Photonics and Quantum Sciences, Heriot-Watt University, Edinburgh, EH14 4AS, UK}
\author{Alexander Arriola}
\affiliation{SUPA, Institute of Photonics and Quantum Sciences, Heriot-Watt University, Edinburgh, EH14 4AS, UK}
\author{ Tim A. Birks}
\affiliation{Department of Physics, University of Bath, Claverton Down, Bath BA2 7AY, UK}
\author{Jeremy R. Allington-Smith}
\affiliation{Department of Physics, University of Durham, South Road, Durham, DH1 3LE, UK}
\author{Robert R. Thomson}
\affiliation{SUPA, Institute of Photonics and Quantum Sciences, Heriot-Watt University, Edinburgh, EH14 4AS, UK}

\begin{abstract}
The spectral resolution of a dispersive astronomical spectrograph is limited by the trade-off between throughput and the width of the entrance slit. Photonic guided-wave transitions have been proposed as a route to bypass this trade-off, by enabling the efficient reformatting of incoherent seeing-limited light collected by the telescope into a linear array of single modes: a pseudo-slit which is highly multimode in one axis but diffraction-limited in the dispersion axis of the spectrograph. It is anticipated that the size of a single-object spectrograph fed with light in this manner would be essentially independent of the telescope aperture size. A further anticipated benefit is that such spectrographs would be free of `modal noise', a phenomenon that occurs in high-resolution multimode fibre-fed spectrographs due to the coherent nature of the telescope point-spread-function (PSF). We address these aspects by integrating a multicore fibre photonic lantern with an ultrafast laser inscribed three-dimensional waveguide interconnect to spatially reformat the modes within the PSF into a diffraction-limited pseudo-slit. Using the CANARY adaptive optics (AO) demonstrator on the William Herschel Telescope, and 1530\,$\pm$\,80\,nm stellar light, the device is found to exhibit a transmission of 47\,--\,53\,\% depending upon the mode of AO correction applied. We also show the advantage of using AO to couple light into such a device by sampling only the core of the CANARY PSF. This result underscores the possibility that a fully-optimised guided-wave device can be used with AO to provide efficient spectroscopy at high spectral resolution.
\end{abstract}

\maketitle

\section{Introduction}\label{intro}

Spectrographs of unprecedented precision will be required to meet the ambitious science goals of modern astronomy, in areas such as Earth-like exoplanet detection via the radial velocity technique \citep{Mayor1995}, and the Sandage test of the real-time rate of expansion of the universe \citep{Sandage1962}. To achieve the required precision, future spectrographs must operate at high spectral resolution (R = $\lambda$/d$\lambda$ $>$  100,000), must be precisely calibrated, and must be exceptionally stable during any given measurement.

The requirement for high precision spectrographs, operating efficiently on larger telescopes over a range of wavelengths, poses a variety of challenges to the design and construction of suitable spectrographs. For example, there tends to be a strong coupling between the size of a telescope and the size of the instruments required to efficiently process the light it captures. This coupling stems from Eq. \ref{Eq1}, where it can be seen that the number of modes that form the Point Spread Function (PSF) of a telescope scales with $(D_{\rm{T}}/4\lambda)^2$ \citep{Harris2012, Spaleniak2013}.

\begin{equation}
\centering
M\approx(\pi\ \theta_{\rm{Focus}} D_{\rm{T}}/4\lambda)^2,
\label{Eq1}
\end{equation}
where $M$ is the number of modes that form the telescope PSF (for each polarisation state), $D_{\rm{T}}$ is the diameter of the telescope and $\lambda$ is the wavelength of the light.  $\theta_{\rm{Focus}}$ is the angular width of the PSF, obtained from a deconvolution of the diffraction-limited and seeing-limited images, but can be approximated as: 
\begin{equation}
\centering
\theta_{\rm{Focus}}\approx \sqrt{(\lambda/D_{\rm{T}})^2 + \theta_{\rm{Seeing}}(\lambda)^2},
\label{Eq2}
\end{equation}
where $\theta_{\rm{Seeing}}(\lambda)$ is the so-called `astronomical seeing' measured as the Full Width at Half Maximum (FWHM) of the long-exposure PSF of the site in radians.

Eq. \ref{Eq1} indicates that the number of modes that form a telescope PSF increases rapidly with increasing telescope aperture, and thus implies that larger telescopes require spectrographs with larger input slit-widths to maintain throughput (of course this is assuming that the F-ratio of the optical input to the spectrograph is held constant). A larger slit-width then results in a larger spectrograph to maintain the required spectral resolution. The larger the instrument, the more expensive it is, and the lower the quality of the optics. The current generation of $\sim$\,8\,--\,10\,m class telescopes already require appropriately scaled spectrographs, in order to obtain the desired resolving power with efficient input coupling, and future $\geq$ 30 m class Extremely Large Telescopes would require even larger instruments \citep{Cunningham2009}.

One approach to address the coupling between telescope size and spectrograph size implied by Eq. \ref{Eq1} is to employ adaptive optics (AO). A perfect AO system would, of course, produce a PSF that was diffraction-limited regardless of the size of the telescope. State-of-the-art extreme AO systems are now able to produce near-diffraction-limited PSF's ($>$ 90 \% Strehl) on 8 m class telescopes at near-IR wavelengths in the H-band, but only over a narrow field of view \citep{Bailey2014, Close2014, Fusco2014, Jovanovic2014, Macintosh2014}. It is also not yet clear how well AO systems will work on the new Extremely Large Telescopes currently under construction; the European Extremely Large Telescope (E-ELT), the Thirty Metre Telescope (TMT) and the Giant Magellan Telescope (GMT), particularly at shorter wavelengths. 

One approach that may complement AO systems, and enable the design of simpler and smaller spectrographs for larger telescopes, is the so-called PIMMS (Photonic Integrated Multimode Micro-Spectrograph) concept \citep{Bland-Hawthorn2010}. The idea behind the PIMMS concept is to use a guided-wave transition, known as a `photonic lantern' \citep{Leon-Saval2005, Bland-Hawthorn2011, Thomson2011, Spaleniak2013, Birks2015}, to efficiently couple the multimode telescope PSF to an array of single modes. These single modes can then be re-arranged (or \emph{reformatted}) into a linear array to form a pseudo-slit that acts as a diffraction-limited single-mode input (along the dispersion axis) to a spectrograph. Importantly, although the length of the slit in the PIMMS concept increases as the number of modes in the PSF increases, the size of the spectrograph in the plane of dispersion is independent of the telescope size.

A further and potentially powerful benefit of the PIMMS concept is its potential to enable fibre-fed spectrographs that are precisely calibrated \citep{Probst2015} and free of modal noise. Modal noise is a phenomenon present in multimode optical-fibre-fed spectrographs, where changes in the modal pattern at the output of the fibre effectively result in variations in the spectrograph linefunction \citep{Lemke2011, McCoy2012}. Interestingly, modal noise is expected to be completely absent in the single mode regime, since the output of the fibre can exhibit only one spatial profile (neglecting polarisation effects), is worst in the two mode regime, and then becomes less severe due to statistical averaging as the number of modes in the fibre increases. If we assume that the fibre is designed to match the PSF, then Eq. \ref{Eq1} would imply that unless we are operating at the diffraction-limit, using single mode fibres, modal noise can become an increasing problem as the wavelength of operation increases, and/or AO systems improve. Modal noise has, for example, now been identified as being a critical issue in GIANO \citep{Iuzzolino2014}, a fibre-fed radial velocity spectrograph intended for operation at between 950 and 2500\,nm on the 3.58\,m Telescopio Nazionale Galileo (TNG). Under 0.6 arcsec seeing, the PSF of the TNG consists of $\sim$75 modes at 950\,nm and just $\sim$12 modes at 2500\,nm. Given these numbers, it is not at all surprising that the performance of GIANO is susceptible to significant modal noise issues.

Previously we have demonstrated that a fully integrated three-dimensional photonic device, which we named a `photonic-dicer' \citep{MacLachlan2016}, could be used to reformat the multimode PSF from the CANARY AO system on the 4.2\,m William Herschel Telescope (WHT) into a diffraction-limited pseudo-slit. The device, which was fully fabricated using an advanced laser manufacturing technique known as Ultrafast Laser Inscription (ULI) \citep{Davis1996, Jovanovic2013}, seamlessly and monolithically integrated a photonic lantern transition with a spatial reformatting section. The device was tested on-sky by feeding the photonic-dicer with H-band (1450\,--\,1610\,nm) stellar light directly from CANARY, and was found to exhibit an on-sky throughput of 20\,\% \citep{Harris2015}, somewhat less than the 65\,\% measured in the laboratory due to sub-optimal input coupling. Although this work proved the feasibility of using ULI fabricated photonic reformatting components for applications in PIMMS-type instruments, for high-precision instruments it will be highly desirable, if not essential, to de-couple the instrument from the telescope slewing, in order to maximise the stability of the instrument. This will almost certainly be achieved using an optical-fibre feed to transport the light from the telescope focal plane to the spectrograph. One potential option to enable this would be to connect the photonic dicer to a standard multimode fibre, but recent work at Macquarie University has suggested that such an approach only induces a new form of modal noise that is due to strongly wavelength dependent coupling losses at the fibre-lantern interface (Cvetojevic et al. in prep).

\begin{figure*}
	\includegraphics[width=2\columnwidth]{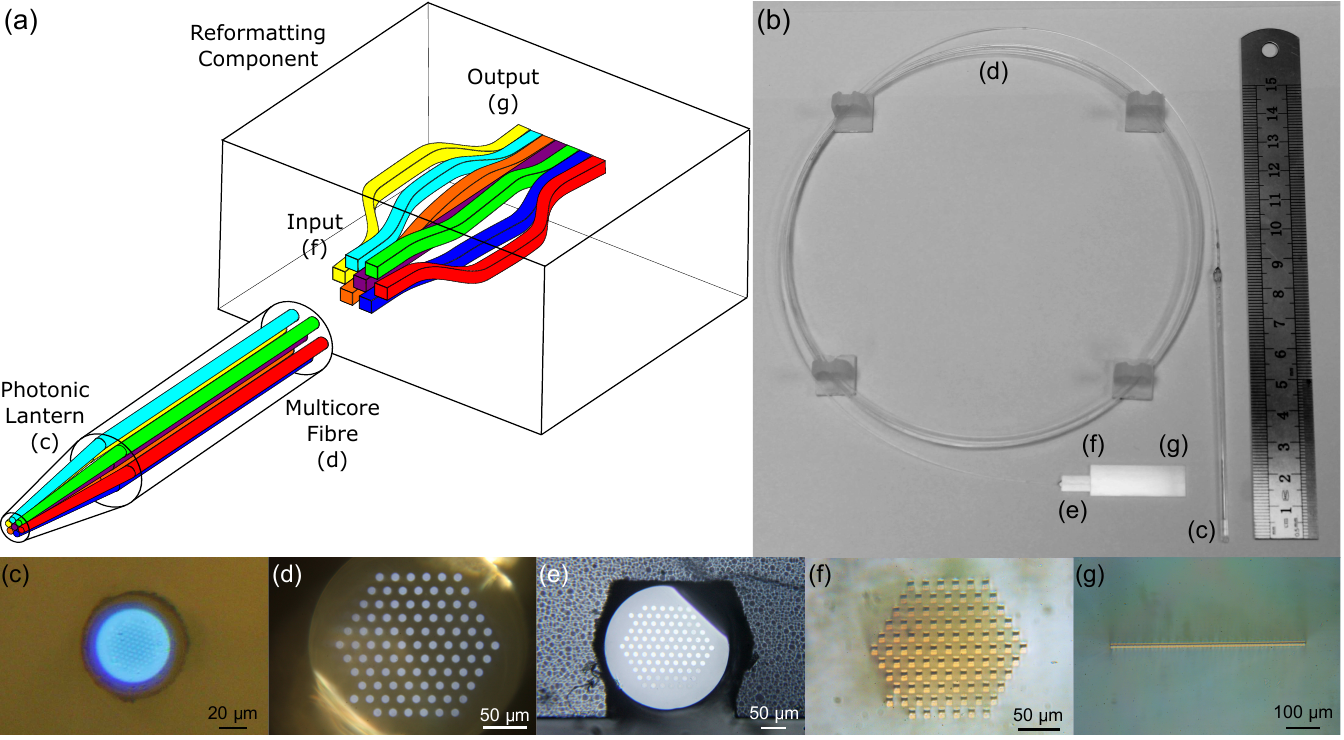}
    \caption{(a) 3-D schematic of a simplified 7 - core hybrid reformatter. The colours differentiate different waveguide paths. (b) Photograph of the hybrid reformatter device. (c) Photonic lantern with multimode input port. (d) Facet of the multicore fibre, with the extra 92nd core visible in the bottom left. (e) Multicore fibre placed in a custom ULI fabricated V-groove. (f) Input facet of the ULI manufactured reformatting component. (g) Pseudo-slit output of the reformatting component.}
    \protect\label{fig:label}
\end{figure*}

With the issues above in mind, an alternative approach is to use a multicore fibre (MCF) photonic lantern to collect the PSF from the telescope, and use the MCF to transport the light from the telescope focal plane to the spectrograph. Such an approach has recently been implemented on the UK Schmidt telescope \citep{Betters2014}, where the two-dimensional array of modes generated by the MCF lantern was then fed directly into a compact spectrograph using the `TIGER' approach \citep{Leon-Saval2012, Betters2013}. Unfortunately, as noted by Betters et al., the lantern used in this work was not correctly matched to the spectral range of the on-sky measurements, and the single-mode cores of the lantern were in fact few-moded at the measurement wavelength. Thus, the full capabilities of implementing photonic-lantern-enabled single-mode spectrographs on-sky, on an astronomical telescope, remain unproven. 

In this paper, we report the successful on-sky application of a `hybrid' photonic-reformatter technology based on an MCF photonic lantern and a ULI fabricated reformatting component \citep{Thomson2012} (Fig. \ref{fig:label}) that reformats an AO-corrected H-band telescope PSF into a diffraction-limited pseudo-slit. The reformatting device presented here exhibited an on-sky throughput of 53\,$\pm$\,4\,\% over a wavelength range of 1530\,$\pm$\,80\,nm, very close to the 65\,\% throughput measured in the laboratory over a wavelength range of 1550\,$\pm$\,20\,nm. We believe that such a hybrid approach, utilising the key capabilities of both MCF photonic lanterns and 3D waveguide technologies, may enable compact high-resolution multimode spectrographs that operate at the diffraction-limit and are free from modal noise.

Section~\ref{reformatter} of the paper describes the design of the MCF lantern and the ULI fabricated component that form the hybrid reformatter. Section~\ref{canary} provides a description of the experimental setup used for the on-sky test. Section~\ref{results} presents the results and analysis of the on-sky testing of both the hybrid reformatter and MCF lantern. Finally Section~\ref{conc} summarises our findings and outlines our planned future work in this field.

\section{Hybrid reformatter design}\label{reformatter}

We aimed to develop a hybrid device that could operate efficiently using the CANARY AO system, an on-sky AO demonstrator system \citep{Myers2008} installed at the 4.2\,m WHT in La Palma. To enable efficient operation, the hybrid reformatter must support at least the same number of modes as the number of modes that form the telescope PSF. The number of modes that form the CANARY PSF was calculated using Eq.~\ref{Eq1} and Eq.~\ref{Eq2}. For our calculations, $D_{\rm{T}}$ is the diameter of the WHT (4.2 m). The central wavelength ($\lambda$) of our planned on-sky measurements was 1530\,nm \citep{Vidal2014}, determined from the overlap between the passband of the H-band filter and the responsivity of the Xenics Xeva-1.7 320 InGaAs camera used as a near-infrared imager for the experiment. Under closed-loop operation (see Section \ref{canary}), a typical AO corrected PSF for the central wavelength in our experimental passband, has a Strehl Ratio of 20 - 30 \% \citep{Gendron2011}. By considering the shortest wavelength within the FWHM of the passband ($\lambda$\,=\,1450\,nm) and using the median seeing at the WHT (0.7\,arcsec) in Eq.~\ref{Eq1}, the number of modes that form the AO corrected PSF is 25, under closed-loop AO correction. This is in contrast to the seeing-limited case, where the number of modes is $\approx$\,60. Given these calculations, the  hybrid reformatter was designed to support in excess of 60 orthogonal modes in order to ensure that efficient input coupling is maintained even under varying atmospheric seeing.

\begin{figure*}
 \includegraphics[width=190mm,trim={5mm 95mm 0 95mm},clip]{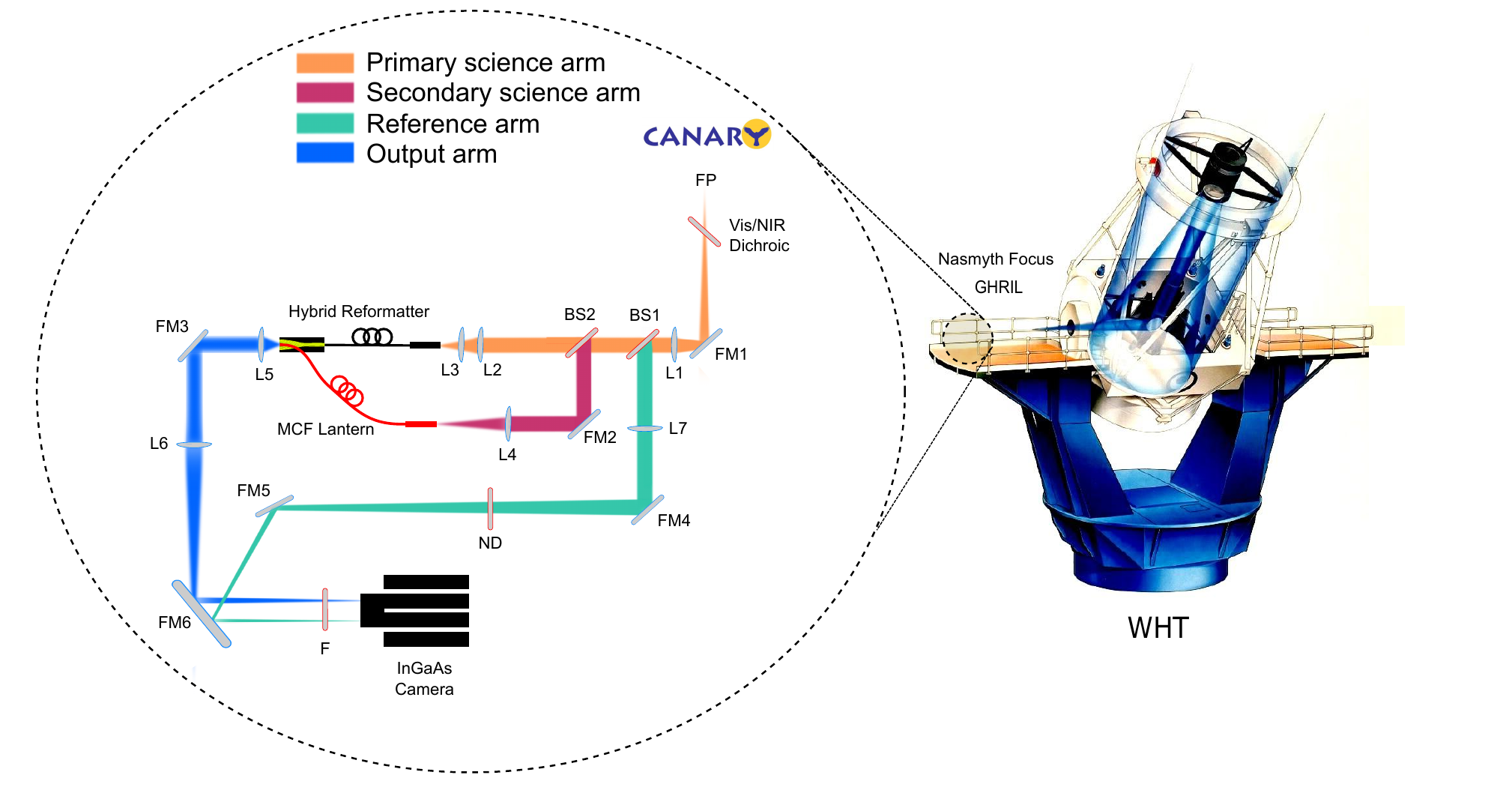}
 \caption{Schematic diagram showing the free-space optical setup used for the on-sky tests. Stellar light collected by the William Herschel Telescope is fed through the CANARY AO system which generates a corrected multimode PSF. A dichroic is used to remove the visible part of the spectrum ($<1000$~nm) from the light beam, which is then collimated using a lens (L1). Using a beamsplitter (BS1), 10\,\% of the collimated beam is fed to a reference path containing a 0.6 absorptive neutral density filter (ND), and re-imaged the PSF onto the camera using L7. The remaining light passes through another beamsplitter (BS2), which reflects 81\,\% to the secondary science arm. The light in this arm is focussed by L4 and allows the core of the PSF to be imaged onto the multimode port of the secondary lantern. The rest of the light is focussed by L2 and L3 onto the hybrid reformatter which samples the whole PSF. The reformatted output from both the hybrid reformatter and secondary MCF lantern are imaged onto the camera using lenses L5 and L6. The fold mirrors (FM1 - FM6) are utilised for beam steering. An \textit{H}-band spectral filter (F) was placed at the entrance to the camera. (WHT image courtesy of the Isaac Newton Group of Telescopes, La Palma).}
 \label{fig2}
\end{figure*}

Figures \ref{fig:label} a \& b present a schematic and photograph of the hybrid reformatter. The fibre optics section of the hybrid reformatter consists of a photonic lantern transition (Fig. \ref{fig:label}c) formed from a multicore fibre (Fig. \ref{fig:label}d). The core arrangement in our 92-core MCF was a centred hexagon of 91 cores (Fig. \ref{fig:label}d) with a single additional core asymmetrically placed at the edge of the pattern as a marker to uniquely identify the orientation of the fibre, facilitating alignment with the ULI reformatter component. The MCF was made from fused silica, with step-index Ge-doped cores, using the standard stack and draw fibre fabrication technique with full fabrication details reported in \citep{MacLachlan2015}. The adiabatic transition for the lantern was made by tapering the MCF in an F-doped capillary as described in \citep{Birks2012, Birks2015}, to form a multimode input port with a core of cross-sectional diameter 43\,$\upmu$m and a numerical aperture of 0.22 (Fig. \ref{fig:label}c). 
The MCF was secured in a custom designed V-groove (Fig. \ref{fig:label}e) in order to facilitate secure connection to the ULI reformatting element. The reformatting element of the hybrid reformatter was designed to match the 92 single-modes generated by the MCF (Fig. \ref{fig:label}f) and reformat them into a diffraction-limited pseudo-slit (Fig. \ref{fig:label}g). For this purpose, the technique of ULI was used to directly inscribe (in the long axis) the reformatter  inside the volume of a 30\,$\times$\,15\,$\times$\,1\,mm substrate of borosilicate glass (SCHOTT AF45) \citep{Meany2013} with full inscription parameters reported in \citep{MacLachlan2015}. At the input end of the reformatter, 92 single-mode waveguides, each 6.2\,$\upmu$m in width, were arranged in a hexagonal geometry with a centre-to-centre separation of 17.6\,$\upmu$m (Fig. \ref{fig:label}f), to match the measured core separation of the manufactured MCF. At the pseudo-slit end of the reformatter, the waveguides were arranged into a one-dimensional array with a centre-to-centre separation of 6.2\,$\upmu$m and an overall length of $\sim$\,570\,$\upmu$m (Fig. \ref{fig:label}g). The MCF lantern and ULI reformatter component were securely bonded with UV cured adhesive to form the hybrid reformatter. As reported in more detail in \citep{MacLachlan2015}, the throughput of the complete hybrid reformatter was tested in the laboratory over a wavelength of 1550\,$\pm$\,20\,nm to be 65\,$\pm$\,2\,\% for incoherent light.

\section{CANARY setup and integration}\label{canary}

As CANARY is designed as an AO demonstrator it can be configured to many different modes of correction. In this case the system was configured to provide closed-loop AO correction using an on-axis natural guide star as a wavefront reference. A dichroic mirror transmits light $>$\,1000\,nm to the experiment with visible wavelengths reflected to a 36 sub-aperture Shack-Hartmann Wavefront Sensor (WFS). A fast-steering mirror and 56-actuator deformable mirror (DM) are driven by the WFS measurements, providing a partially corrected PSF at a wavelength of 1500\,nm. A basic integrator feedback controller with a closed-loop gain of 0.3 was used to calculate the DM commands, with the WFS positioned behind the DM measuring the residual wavefront error after correction.

\begin{figure*}
	\centering
	 \includegraphics[width=170mm]{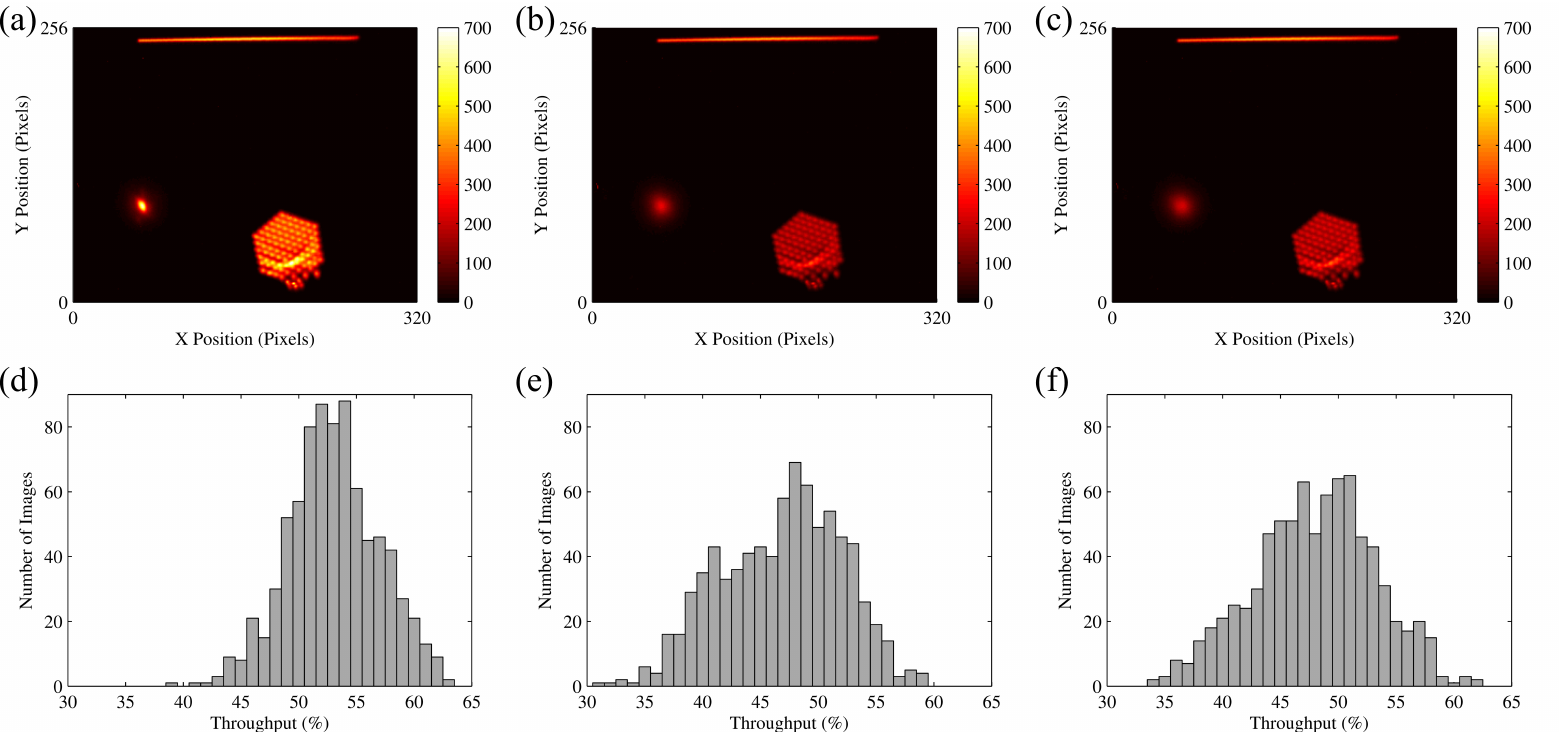}
 \caption{Colour map images of the camera sensor for averaged, reduced data showing the output of the hybrid reformatter (along the very top), the secondary lantern (lower right) and the telescope PSF (lower left) from the reference path for (a) closed-loop, (b) tip-tilt only and (c) open-loop modes of AO correction. Note: The camera has $30\,\upmu$m pixels and the secondary lantern output is distorted due to an imperfect cleave. Bottom row: Histogram plots showing percentage transmission distribution over the number of images acquired for (d) closed-loop, (e) tip-tilt and (f) open-loop AO modes. Hot pixel removal and background correction algorithms have been applied prior to evaluating the transmission for each image frame acquired.}
	\protect\label{fig3}
\end{figure*}

Alignment and optimisation of the experimental system (Fig.~\ref{fig2}) was performed during the day using a range of sources available in the CANARY setup. In our case a 1550 nm laser beam coupled into a single-mode fibre was moved into the CANARY input focal plane, passed through the entire CANARY optical train, and the PSF re-imaged onto the multimode input at the photonic-lantern end of the hybrid reformatter. The surface shape of the DM was modified by engaging the AO-correction loop and artificially applying static offset terms to the measured WFS signal. These offsets were automatically adjusted using the Nelder-Mead simplex method \citep{NelderMead1965} to alter the PSF shape at the hybrid reformatter input and maximise the detected flux from its pseudo-slit output. The optimum measured wavefront values and corresponding DM shape were recorded and used as the correction reference for on-sky tests.

To investigate the effect of different degrees of AO correction on the hybrid reformatter, CANARY was operated in three modes. Closed-loop mode provides the maximum degree of correction achievable, where both tip-tilt and higher-order wavefront aberrations were corrected with an update rate of 150\,Hz. In tip-tilt mode the high-order AO correction of the optimised PSF shape is removed by reducing the integrator loop gain to a low value (typically 0.001), with the PSF location simply stabilised in real-time. Open-loop mode applies the minimum correction of the three modes of operation by additionally reducing the tip-tilt correction gain. This allowed the PSF to remain in the reference location for optimum coupling, but without high temporal frequency correction.

As shown in Fig.~\ref{fig2}, the on-sky experimental setup comprised two science arms and a separate reference arm. A beamsplitter (BS1) was used to direct $\sim$\,10 \% of the light within the CANARY PSF to the reference arm, where a lens (L7) formed an image of the PSF onto the InGaAs camera. The reference arm enabled the shape and photon flux within the PSF to be `self-referenced' during observation, allowing the transmission of the hybrid reformatter to be measured regardless of variations in the PSF. A 0.6 absorptive neutral density filter (ND) was placed in this arm to ensure that the camera pixels did not saturate during data acquisition. An \textit{H}-band spectral filter (F) was mounted at the entrance to the camera. A second beamsplitter (BS2) was introduced to direct $\sim$\,81\,\% of the remaining light into the secondary science arm where it was injected into the multimode end of a second MCF photonic lantern (which we will call the secondary lantern) using lens L4. The final  $\sim$\ 9 \% of the light in the primary science arm was coupled into the hybrid reformatter using lenses L2 and L3. The output ends of the secondary lantern and hybrid reformatter were adjacently positioned in the vertical axis such that both could be imaged \emph{via} lenses L5 \& L6 onto the camera simultaneously. The detector response of the InGaAs camera was measured in the laboratory using a 1550 nm light emitting diode with a calibrated integrating sphere detector and found to be linear to within $\ll$\,1\,\% over the signal levels used throughout the experiments. 

\begin{figure*}
	\centering
	 \includegraphics[width=170mm]{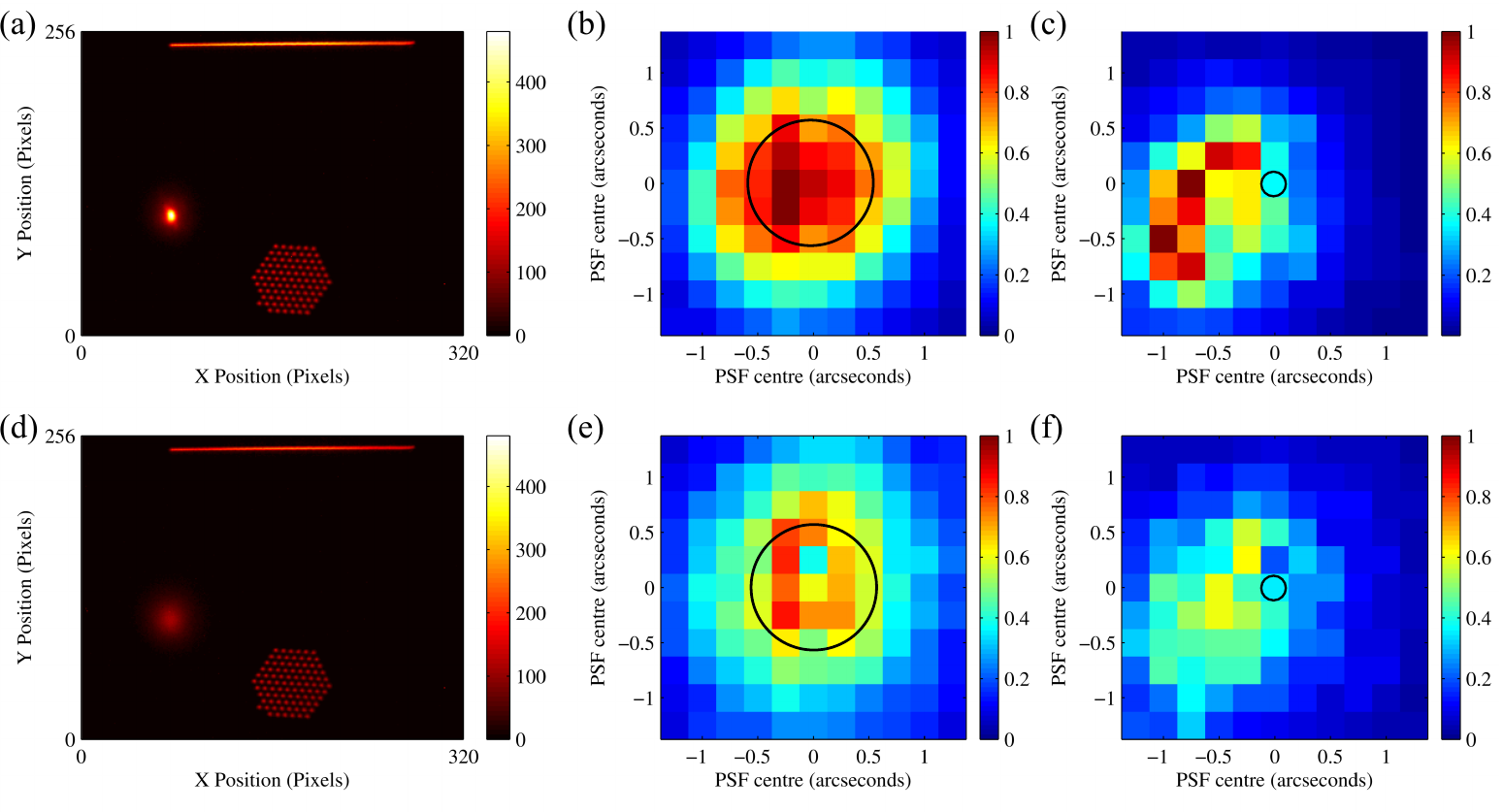}
\caption{Dither plots for the night of 2014 October 12. Top row shows closed-loop operation, and bottom row tip-tilt operation. Left column shows averaged reduced data obtained from the camera. Middle column images show the position of the dithered PSF, with the colour bar being normalised flux for the slit of the hybrid reformatter. The right column images show the position of the dithered PSF, with the colour bar being normalised flux for the secondary photonic lantern. The black circles represent the angular size of the input fibre in each science arm}
	\protect\label{fig4}
\end{figure*}

CANARY provides an $\approx$\,F/11 beam with a plate scale value of 4.54\,arcsec/mm, but this is obviously altered by the input coupling optics in the two arms (L2 to L4 in Fig.~\ref{fig2}). The scaling of each arm was measured by taking images at the focus of each arm, while translating an IR source at another known focal plane within CANARY.  The primary science arm was designed to couple the entire PSF, with an F/2.29\,$\pm$\,0.28 beam and a plate scale of 21.8\,$\pm$\,1.34\,arcsec/mm. This means the hybrid reformatter input had an angular size of 1.1\,$\pm$\,0.07\,arcsec. The secondary science arm was designed to couple the core of the PSF, with an F/11.07\,$\pm$\,0.32 beam and a plate scale of 4.57\,$\pm$\,0.13\,arcsec/mm. This means the secondary MCF lantern input had an angular size of 0.23\,$\pm$\,0.01\,arcsec.

For each mode of AO operation, multiple datasets were acquired, where each dataset comprised 100 near-infrared images. The camera exposure time for each image was 400\,ms. The background noise floor was determined through dark exposures and  periodically acquiring datasets of sky background images. After acquiring an adequate number of datasets, the hybrid reformatter and secondary lantern were removed from the primary science arm and the lens L5 was translated towards lens L3, such that the PSF could be imaged directly onto the camera through the primary science arm. Additional datasets of 100 images were acquired in order to calibrate the instrument for the difference in the integrated power between the reference and primary science arms. This allowed the throughput of the hybrid reformatter to be calibrated using the star itself.

\section{Results}\label{results}

The hybrid reformatter and the secondary lantern were tested on-sky at the WHT on 2014 October 11 \& 12, with all data acquired between 20:30 and 22:45 GMT. The star selected for observation was TYC 3156-2223-1 (gamma Cygni) from the Tycho 2 catalogue \citep{Hog2000}, a $1^{\rm{st}}$ magnitude star in the astronomical \textit{H}-band. The astronomical seeing values, as measured using an on-site monitor \citep{OMahoney2003}, varied between 0.5\,and\,0.8\,arcsec over the course of the measurements, representative of median seeing for the telescope site.

Fig.~\ref{fig3} shows the near-infrared images for each AO mode and results from the corresponding throughput evaluation using data acquired on the night of 2014 October 11. The images presented here (Fig.~\ref{fig3} top) have been averaged over 100 frames and show the pseudo-slit and secondary lantern outputs that were imaged onto the camera using the primary science arm, as well as the CANARY PSF, which was imaged using the reference arm (see Fig.~\ref{fig2}). The imperfect cleave forming the multicore end of the secondary lantern can also be seen in the images. In the case of closed-loop operation with full AO correction, the transmission of the hybrid reformatter was measured to be 53\,$\pm$\,4\,\%. With reduced degree of atmospheric correction, the device transmission was measured to be 47\,$\pm$\,5\,\% and 48\,$\pm$\,5\,\% in the case of tip-tilt and open-loop operation respectively. Histograms of the transmission data obtained for the respective AO modes are shown in Fig.~\ref{fig3} bottom.

For experiments conducted on the night of 2014 October 12, the multicore end of the secondary lantern was re-cleaved to produce an improved image on the camera. As the experimental design did not allow for easy calibration of a reference PSF image to be obtained in the secondary science arm, the absolute throughput of the secondary lantern was not determined. Nonetheless, comparative measurements between the closed-loop and tip-tilt modes of AO operation were performed. These were taken in the same manner as the throughput measurements, with the calculated closed-loop and tip-tilt flux divided to provide a ratio of the two. These measurements revealed an averaged flux ratio of 2:1, which is much lower than that expected, according to the area of the sampled PSF in both cases. To examine this discrepancy, an on-sky dither was performed whereby the tip-tilt mirror in the CANARY setup was used to move the PSF by $\pm$ 1.25 arcsec relative to the inputs of the hybrid reformatter and the secondary lantern, which have an effective angular size of 1.1 and 0.23 arcsec respectively.

Fig. \ref{fig4} shows the results of the dither experiments for the respective devices, with each square representing the total flux of the tested device for each PSF position. The centre of each plot shows the location (in arcseconds) of the centre of the PSF as set on the bench, with the black circles representing the respective angular size of the device input. The dither plots clearly show that the actual area of maximum flux does not overlap with the expected area. This mismatch can be attributed to atmospheric refraction causing a chromatic shift between the corrected (visible guide star) path and the science (near-infrared) path and will be corrected in future experimentation.

It can also be seen from Fig. \ref{fig4} middle column, that the hybrid reformatter is relatively insensitive to dither position in both closed-loop and tip-tilt operating modes, with closed-loop providing approximately 35\,\% improvement over tip-tilt correction at the optimum input position. The secondary lantern (Fig. \ref{fig4} right column) is also insensitive in tip-tilt mode, though highly sensitive in closed-loop mode and demonstrates approximately 58\,\% improvement in throughput between closed-loop and tip-tilt correction. This shows that if lanterns are designed for few modes they will need to be carefully matched to the AO system they sit behind. It should be noted that the dithers did not incorporate the simplexing routine that was applied during the throughput demonstration of the previous night, and so were not fully optimised for off-axis positions.

\section{Conclusions}\label{conc}
We have presented results from the on-sky tests of a photonic guided-wave device that spatially reformats an AO corrected telescope PSF into a diffraction-limited pseudo-slit. This hybrid reformatter integrates a low-loss MCF photonic lantern that adiabatically converts the modes within the PSF into a two-dimensional array of single-modes, with a ULI fabricated reformatter component that spatially rearranges the single modes into a pseudo-slit waveguide that is single mode in the dispersion axis. Our aim is that devices based on this concept will convert any modal-noise into amplitude and phase variations orthogonal to the dispersion axis of an instrument, and could enable high-resolution, high-stability spectrographs operating on large telescopes, or in situations where modal noise is a severe issue. The calibration tools and controls required to optimise coupling to an AO system were developed and the device performance tested for a range of modes of AO operation. The maximum on-sky throughput achieved was 53\,$\pm$\,4\,\%  while the in-laboratory throughput was measured to be 65\,$\pm$\,2\,\%. This performance demonstrates a significant improvement over that achieved in our earlier demonstration of the fully integrated ULI reformatter (the photonic dicer).

To conclude, we believe the results presented here highlight the potential of efficiently implementing guided-wave reformatting devices in astronomy, specifically for coupling multimode light to spectrographs operating at the diffraction-limit. This approach may have important applications in many areas, such as radial velocity measurements and exoplanet atmospheric spectroscopy.

\section*{Acknowledgments}
R.R.T. gratefully acknowledges funding from the STFC in the form of an STFC Advanced Fellowship (ST/H005595/1), from the Royal Society (RG110551) and from Renishaw plc.  D.G.M. is supported by an EPSRC Ph.D studentship. R.R.T. and T.A.B. acknowledge funding from the STFC-PRD scheme (ST/K00235X/1). R.J.H. and J.R.A-S. gratefully acknowledge support from the Science and Technology Facilities Council (STFC) in the form of a Ph.D studentship (ST/I505656/1) and grant (ST/K000861/1).  R.R.T., T.A.B. and J. R. A-S thank the European Union for funding via the OPTICON Research Infrastructure for Optical/IR astronomy (EU-FP7 226604). CANARY was supported by Agence Nationale de la Recherche program 06-BLAN-0191, CNRS / INSU, Observatoire de Paris, and Universit{\'e} Paris Diderot Paris 7 in France, STFC (ST/K003569/1 and ST/I002781/1), University of Durham in UK and European Commission Framework Programme 7 (E-ELT Preparation Infrastructure Grant 211257 and OPTICON Research Infrastructures Grant 226604). Raw experimental data will be made available on the Heriot-Watt PURE system.

\end{document}